\documentstyle[twocolumn,aps]{revtex}
\begin{document}

\draft
\title{Black holes in the Brans-Dicke-Maxwell theory}
\author{Rong-Gen Cai $^{a,b}$ and Y. S. Myung $^c$}
\address{ $^a$ CCAST (World Laboratory), P. O. Box 8730, Beijing
100080, China \\
$^b$ Institute of Theoretical Physics, Academia Sinica, P. O.
Box 2735, Beijing 100080, China  \\
$^c$ Department of Physics, Inje University, Kimhae 621-749, Korea}
\maketitle

\begin{abstract}
The black hole solutions in the higher dimensional 
Brans-Dicke-Maxwell theory are investigated.
We find that the presence of the nontrivial scalar field  depends
on the spacetime dimensions $(D)$. When $D=4$, the 
solution corresponds to the Reissner-Nordstr\"{o}m black hole with a  
constant scalar field. In higher dimensions $(D>4)$, one finds the
charged black hole solutions with the nontrivial scalar field. The 
thermal properties of the charged black holes are discussed and the 
reason why the nontrivial scalar field exists are explained.
Also the solutions for  higher dimensional Brans-Dicke theory are  
given for comparison.

\end{abstract}

As is well known, differing from  general relativity with the  metric,
 the Brans-Dicke (BD) theory [1]
describes the gravitation in terms of  the metric as well as a
scalar field. Due to the scalar field, the BD
theory and general relativity must have distinctions in some domains,
 although they can be in agreement under the
 post-Newtonian approximation. In recent years, much attention has
been drawn into the BD theory, in particular, in the strong
 field domains.
A strong field appeared in the early universe.  La and Steinhardt [2]
 have shown that the BD theory seems to be better than the Einstein
 theory of gravity for solving the ``graceful exit'' problem in the 
inflation model. This is  because the scalar field in the BD theory 
provided a natural termination
of the inflationary era via bubble nucleation without the need for 
finely tuned cosmological parameters.

The other example  comes from the black holes in the BD theory.
 More recently, many authors have investigated the gravitational 
 collapse and black hole formation in the BD theory [3,4,5,6,7].
  It turned out that the dynamic
scalar field in the BD theory plays an important role 
in the process of collapse and critical phenomenon.
 Hawking [8] proved first that in the
 four dimensional vacuum BD theory, the  black hole solution is just
the Schwarzschild solution with a trivial constant scalar field 
(hereafter the black
holes in this paper mean the static, asymptotically flat, and
spherically symmetric  solutions with horizon). Further the stability of 
 black holes in the BD theory has  been investigated in Ref. [9]. On the
 other hand, the vacuum BD
theory can  be transformed into the Einstein-massless scalar theory by 
using a conformal transformation.
In Ref. [10], the  solution to Einstein-massless scalar  equations
 was given. Although this solution has an asymptotically flat 
region and the  scalar field approaches zero at
spacelike infinity, it exhibits a naked singularity. When the scalar
field is constant, the solution  reduces to the Schwarzschild
case. It is also noted that the  black hole solution in
 the vacuum BD theory corresponds to the Schwarzschild solution 
 with a constant scalar field. This can also be confirmed from the
no scalar-hair theorem by Bekenstein [11] and Saa [12].

On the other hand, it is well known that the black hole solution to
 Einstein-Maxwell equations is the Reissner-Nordstr\"{o}m solution. In 
higher dimensions, its solution  can be regarded as a simple dimensional
 generalization of Reissner-Nordstr\"{o}m solution [13]. In order
 to investigate the distinctions between the BD and   Einstein theories,
 it is important to see whether the black hole solution in the
 Brans-Dicke-Maxwell theory belongs to the Reissner-Nordstr\"{o}m solution
or its trivially dimensional extension. We find that the $D=4$
   black hole solution in the BD-Maxwell theory belongs to
the Reissner-Nordstr\"{o}m solution with a constant scalar field. In 
higher dimensions $(D>4)$, however, one obtains  the black hole solutions
with the nontrivial scalar. This is because the
 stress-energy tensor of  Maxwell field is not traceless in
the higher dimensions and the action of  Maxwell
field is not  invariant under conformal transformations.
Accordingly, the Maxwell field can be regarded as the source of the
scalar field in the BD theory. The main purpose of this paper is
to report this result.

In the $D(\ge 4$) dimensions, the action of the
Brans-Dicke-Maxwell theory is given by
\begin{equation}
I=\frac{1}{16\pi}\int d^Dx\sqrt{-g}(\phi R-\frac{\omega}{\phi}
g^{\mu\nu}\nabla_{\mu}\phi \nabla_{\nu}\phi-F_{\mu\nu}F^{\mu\nu}),
\end{equation}
where $R$ is the scalar curvature, $F_{\mu\nu}$ is the Maxwell field,
$\omega$ is the coupling constant, and the $\phi$ denotes the BD scalar
 with the dimensions $G^{-1}$. Here $G$ is a $D$-dimensional
Newtonian  constant. In this paper, we choose  units such
that $c=G=1$.  In this BD frame, test particles have constant rest
 mass and
move along the geodesics. That is, matter fields are coupled to gravity
 only via the metric, and do not interact with the scalar $\phi$.
So we introduce the Maxwell kinetic term as in (1). Varying  (1) yields  
equations of motion,
\begin{eqnarray}
\phi G_{\mu\nu}&\equiv & \phi (R_{\mu\nu}-\frac{1}{2}g_{\mu\nu}R)
\nonumber\\
 &=&\frac{\omega}{\phi}\left [\nabla _{\mu}\phi \nabla _{\nu}\phi -
\frac{1}{2}g_{\mu\nu}(\nabla \phi )^2\right ] \nonumber\\
&+&2(F_{\mu}^{\, \lambda}F_{\nu\lambda}-\frac{1}{4}g_{\mu\nu}F^2)
\nonumber\\
&+&\nabla _{\mu}\nabla _{\nu}\phi -g_{\mu\nu}\nabla ^2 \phi,\\
0&=&\nabla _{\mu}(F^{\mu\nu}),\\
\nabla ^2 \phi &=&-\frac{d-1}{2[(d+1)\omega +(d+2)]}F^2,
\end{eqnarray}
where $G_{\mu\nu}$ is the Einstein tensor, $\nabla$ represents the
covariant differentiation in the spacetime metric $g_{\mu\nu}$ and
$d=D-3$. Solving (2)-(4) directly is a non-trivial task because the 
right hand side of (2) includes the second derivatives  of the
scalar. We can
remove this difficulty by a conformal transformation.

Considering a conformal transformation
\begin{equation}
g_{\mu\nu}=\Omega ^2 \bar{g}_{\mu\nu}
\end{equation}
 with
\begin{equation}
\Omega ^{-(d+1)}=\phi
\end{equation}
and
\begin{equation}
\bar{\phi}=\sqrt{2a}\int ^{\phi} \frac{d\phi}{\phi}=
\sqrt{2a}\ln\phi,~~~ a=\frac{d+2}{d+1}+\omega,
\end{equation}
the BD-Maxwell theory (1) can be transformed into
the  Einstein-Maxwell theory with a minimally coupled scalar field 
$(\bar{\phi})$
\begin{equation}
\bar{I}=\frac{1}{16\pi}\int d^Dx\sqrt{-\bar{g}}[\bar{R}-\frac{1}{2}
(\bar{\nabla}\bar{\phi} )^2 -e^{-b\bar{\phi}}\bar{F}^2],
\end{equation}
where
\begin{equation}
b=\frac{d-1}{d+1}\frac{1}{\sqrt{2a}},
\end{equation}
 $\bar{R}$ and $\bar{\nabla}$ are the scalar curvature and covariant
 differentiation in the new metric $\bar{g}_{\mu\nu}$, respectively.
 Here a few points should be stressed. First of all,  (7) implies 
 $a>0$ ($ \omega >-\frac{d+2}{d+1}$), and one has
$\bar{\phi}=0$
at spacelike infinity. Second, the action remains unchanged under the
conformal transformation ($\bar{I}$ and $I$ give us  a difference of 
 surface term associated with the scalar field). This point plays an
important role in dealing with physical quantities
 between the two frames. Third, we stress that the BD theory (1) is 
 only mathematically equivalent to the theory (8). In the Einstein
frame (8), the test particle will take variable rest mass with
spacetime and  move no longer  along the geodesics. Note 
that there exists a coupling between  Maxwell field and 
scalar field.  Finally, it is worth noting that $b=0$ when $D=4$. 
In four dimensions the Maxwell field is decoupled from the scalar
 field in the Einstein frame and further in the BD frame,  it cannot
  be considered as the source of the scalar field $\phi$ [see (4)].
Hence it turns out that $D=4$ is a special case in the BD-Maxwell 
theory.

Varying the action (8), we can obtain equations of motion
\begin{eqnarray}
\bar{G}_{\mu\nu}&\equiv &\bar{R}_{\mu\nu}-\frac{1}{2}
\bar{g}_{\mu\nu}\bar{R} \nonumber\\
&=&\frac{1}{2}\bar{\nabla} _{\mu}\bar\phi \bar{\nabla} 
_{\nu}\bar{\phi}-\frac{1}{4}\bar{g}_{\mu\nu}(\bar{\nabla}
 \bar{\phi})^2 \nonumber \\
&+&2e^{-b\bar{\phi}}(\bar{F}_{\mu}^{\, 
\lambda}\bar{F}_{\nu\lambda}-\frac{1}{2}\bar{g}_{\mu\nu}\bar{F} ^2),\\
\bar{\nabla} ^2\bar{\phi}&=&-be^{-b\bar{\phi}}\bar{F}^2,\\
0&=&\bar{\nabla} _{\mu}(e^{-b\bar{\phi}}\bar{F}^{\mu\nu}).
\end{eqnarray}
Comparing  (2)-(4) with  (10)-(12), one finds that if
($\bar{g}_{\mu\nu}$, $\bar{\phi}$, $\bar{F}_{\mu\nu}$) is the 
 solution to (10)-(12), then
\begin{equation}
 (g_{\mu\nu}, \phi , F_{\mu\nu})=(e^{-\frac{2}{(d+1)\sqrt{2a}}
 \bar{\phi}}
\bar{g}_{\mu\nu}, e^{\frac{1}{\sqrt{2a}}\bar{\phi}}, \bar{F}_{\mu\nu})
\end{equation}
is  the  solution of  (2)-(4). In order to demonstrate clearly,
 let us consider first the
 absence of the Maxwell field. In the four dimensions,
Brans [14] has constructed  the static
solutions in the Brans-Dicke frame. Here we will provide the solutions 
of higher dimensional BD theory by the conformal transformation. 
In the absence of Maxwell field, 
Equations (10)-(12) have the following solution with isotropic
 coordinates [10]:
\begin{eqnarray}
d\bar{s}^2&=&-e^{f}dt^2+e^{-h}(dr^2 +r^2d\Omega ^2_{d+1}), \\
\bar{\phi}&=&\left [\frac{2(d+1)}{d}(1-\gamma ^2)\right ]^{1/2}
\ln\frac{r^d-r_o^d}{r^d+r_o^d},
\end{eqnarray}
where
\begin{eqnarray}
e^f&=&\left [ \frac{r^d-r_o^d}{r^d+r_o^d}\right ]^{2\gamma},\\
e^{-h}&=&\left [ 1-\frac{r_o^{2d}}{r^{2d}}\right ]^{2/d}\left [
\frac{r^d
-r_o^d}{r^d+r_o^d}\right ] ^{-2\gamma/d}.
\end{eqnarray}
Here $\gamma$ and $r_o$ are two integration constants. Obviously, when
 $r\rightarrow \infty$, one has $f\rightarrow 0$, $h\rightarrow 0$,
 and $\bar{\phi}\rightarrow 0$. Therefore the spacetime (14) has
  asymptotically flat region and the scalar field $\bar{\phi}$ 
vanishes at
  spacelike infinity. From (15) it follows $0\le \gamma ^2\le 1$.
When $\gamma \in [-1,0)$, however, the solution has a ``negative''
Arnowitt-Deser-Misner (ADM) mass [10]. In particular, for
$\gamma =-1$ the solution is the D-dimensional Schwarzschild solution
 with a negative mass,
 which describes a naked singularity spacetime. In order for the
solution (14) to be a physical solution, we confine $\gamma$ to $0\le
 \gamma \le 
1$. The special example is $\gamma=1$.
In this case, the scalar field vanishes.  Equation (14) describes a
$D$-dimensional Schwarzschild black hole geometry with mass $M=2r_o^d$. 
In other cases, (14) describes  spacetime with a naked singularity,
whose singular point is  $r=r_o$. To observe it explicitly, let us
calculate the scalar curvature of (14).
 This leads to
\begin{equation}
\bar{R}=\frac{4d(d+1)r_o^{2d}(1-\gamma ^2)r^{2(d+1)}}{(r^d
+r_o^d)^{2(d+1+\gamma )/d}(r^d-r_o^d)^{2(d+1-\gamma)/d}}.
\end{equation}
From (18) it is shown  that in the cases of $\gamma ^2 \ne 1$, 
spacetime (14) has a naked scalar curvature singularity at $r=r_o$, 
which can not be removed by coordinate transformations. Therefore, in
 the higher dimensional ($ D\ge 4$) Einstein-minimally coupling massless
 scalar field system, the only black hole solution is D-dimensional 
Schwarzschild solution with a trivial scalar field.

In the $D$-dimensional vacuum BD theory, using (13), we can  obtain
its solution,
\begin{eqnarray}
ds^2&=&\Omega ^2d\bar{s}^2=\left (\frac{r^d+r_o^d}{r^d-r_o^d}
\right )^{\frac{2}{d+1}\left [\frac{(d+1)(1-\gamma ^2)}{ad}
\right ]^{1/2}} d\bar{s}^2,\\
\phi &=&\left (\frac{r^d-r_o^d}{r^d+r_o^d}\right )^{\left [\frac{(d
+1)(1-\gamma ^2)}{ad}\right ]^{1/2}},
\end{eqnarray}
where $d\bar{s}^2$ is given by Eq. (14).  It is easy to show that the
 solution (19) has asymptotically flat space and the point $r=r_o$ 
 corresponds to
 a naked singularity still. This can be found from calculating
  the scalar curvature of the solution (19) through the relation,
\begin{eqnarray}
R&=&\Omega ^{-2}\bar{R}-2(d+2)\Omega ^{-3}\bar{g}^{\mu\nu}
 \bar{\nabla} _{\mu}\bar{\nabla} _{\nu}\Omega    \nonumber\\
&-&(d+2)(d-1)\Omega ^{-4}\bar{g}^{\mu\nu}\bar{\nabla}_{\mu} \Omega
\bar{\nabla} _{\nu}\Omega .
\end{eqnarray}
Again, when $\gamma=1$, the solution (19) is reduced to the
 D-dimensional Schwarzschild solution with the constant scalar
  field ($\phi =1$). In that
case, the BD theory degenerates into  the Einstein theory of gravitation.
From (15) we find $\bar{\phi}\le 0$ (because of
$r_o \ge 0$). So the scalar  $\phi$ in the BD theory belongs to the
 region $\phi \in (0, 1]$. When the Maxwell field is absent, however,
the action (8) and equations of motion (10)-(12) remain unchanged under
 the transformation: $\bar{\phi}\rightarrow -\bar{\phi}$. Thus, we can
obtain  another  solution of the vacuum BD theory,
\begin{eqnarray}
ds^2&=&\left (\frac{r_d-r_o^d}{r^d+r_o^d}\right )^{\frac{2}{d+1}
\left [\frac{(d+1)(1-\gamma ^2)}{ad}\right ]^{1/2}}d\bar{s}^2,\\
\phi &=&\left (\frac{r_d+r_o^d}{r^d-r_o^d}\right )^{\left [\frac{
(d+1)(1
-\gamma ^2)}{ad}\right ]^{1/2}},
\end{eqnarray}
where $d\bar{s}^2$ is still given by Eq. (14). In this case, the scalar
 field $\phi$ takes values in the region $ [1,\infty)$. But the spacetime
(22) has  still asymptotically flat  region and the point $r=r_o$ is a 
curvature singularity unless $\gamma=1$. When $\gamma=1$, the scalar
field is a constant and the solution (22) is the $D$-dimensional
 Schwarzschild solution. Thus, we emphasized again that the  black hole
solution of the
 vacuum BD theory is the Schwarzschild solution with a constant scalar
 field in  higher dimensions.

We now turn to the charged case.  Introducing the Maxwell field,
 the situation is changed significantly. Consulting with the
 conformal transformation (5)-(7), instead of (2)-(4),  
equations (10)-(12) can be used for looking for the solutions.
 The  black hole solutions for the actions similar to  (8) have been
 found in Refs. [15,16]. Considering the dual
form of the black
holes given in Ref. [16], we obtain the black hole
solutions of (10)-(12),
\begin{eqnarray}
d\bar{s}^2&=&-A^2dt^2+B^2dr^2+C^2d\Omega ^2_{d+1},\\
e^{b\bar{\phi}}&=&\left [1-\left(\frac{r_-}{r}\right)^d
\right]^{\alpha d},\\
\bar{F}_{tr}&=&\frac{Q}{r^{d+1}},
\end{eqnarray}
where
\begin{eqnarray}
A^2(r)&=&\left [1-\left(\frac{r_+}{r}\right )^d\right ]\left [1-
\left(\frac{r_-}{r}\right )^d\right ]^{1-\alpha d},\\
B^2(r)&=&\left [1-\left(\frac{r_+}{r}\right )^d\right ]^{-1}\left [1-
\left(\frac{r_-}{r}\right )^d\right ]^{\alpha -1}\\
C^2(r)&=&r^2\left [1-\left(\frac{r_-}{r}\right )^d\right ]^{\alpha},\\
\alpha &=&\frac{2b^2(d+1)}{d(2d+b^2(d+1))}.
\end{eqnarray}
Here $Q$, $r_+$ and $r_-$ are integration constants. According to the 
Gauss theorem, the electric charge  is
\begin{equation}
q=\frac{1}{4\pi}\int _{r\rightarrow \infty}\bar{F}_{tr}\sqrt{-
\bar{g}}d^{d+1}x
=\frac{A_{d+1}}{4\pi}Q,
\end{equation}
where $A_{d+1}$ is the volume of the (d+1)-dimensional unit sphere.
The constant $Q$ is related to the constants $r_+$ and $r_-$,
\begin{equation}
Q^2=\frac{\alpha d^3(r_+r_-)^d}{2b^2}.
\end{equation}
From (25) we see that the scalar field is bounded everywhere, except
 at $r=0$. When $\alpha =0$, the solution (24)-(26) reduces to a
 D-dimensional Reissner-Nordstr\"{o}m solution with the vanishing 
scalar field. Thus, the constants $r_+$ and $r_-$ can be interpreted
 in terms of the outer horizon and inner horizon (assuming
 $r_+\ge r_-$).
 But, for a general $\alpha $, the point $r_-$ is a scalar
  curvature singularity.
 This can be seen from the scalar curvature of
 the solution (24),
\begin{eqnarray}
\bar{R}&=&\frac{\alpha ^2 d^4 r_-^{2d}}{2b^2r^{2d+2}}
\left [1-\left (\frac{r_+}{r}\right )^d\right ]\left [1
-\left (\frac {r_-}{r}\right
 )^d\right ]^{-(\alpha +1)} \nonumber\\
&-&\frac{4Q^2}{r^{2d+2}}\left [1-\left (\frac{r_-}{r}
\right )^d\right ]^{-\alpha},
\end{eqnarray}
which  diverges at $r=r_-$ unless $\alpha =0$. This also confirms that 
the inner horizon of D-dimensional Reissner-Nordstr\"{o}m black holes is
instable. In the our case, due to the appearance of the scalar field, 
the  inner horizon is converted into a scalar curvature singularity.

 With  the Euclidean action method [17, 18], we obtain
 the ADM mass ($\bar{M}$), Hawking temperature ($\bar{T}$), and
  the entropy
($\bar{S}$) of the black hole solution,
\begin{eqnarray}
\bar{M}&=&\frac{A_{d+1}}{16\pi}(d+1)[r_+^d+(1-\alpha -\alpha d)r_-^d],
\\
\bar{T}&=&\frac{d}{4\pi r_+}\left [1-\left (\frac{r_-}{r_+}\right )^d
\right ]^{1-\alpha (d+1)/2},\\
\bar{S}&=&\frac{1}{4} \bar{\Sigma} =\frac{A_{d+1}r_+^{d+1}}{4}\left [1-
\left (\frac{r_-}{r_+}\right )^d\right ]^{\alpha (d+1)/2},
\end{eqnarray}
 where $\bar{\Sigma}$ is the horizon area of black hole (24).
With (13), one finds the charged black hole
solution in the Brans-Dicke-Maxwell theory (frame),
\begin{eqnarray}
ds^2&=&\Omega ^2 d\bar{s}^2=\left [1-\left (\frac{r_-}{r}\right )^d
\right ]^{-2\alpha d/(d-1)}d\bar{s}^2,\\
\phi &=&e^{\frac{1}{\sqrt{2a}}\bar{\phi}}=\left [1-
\left (\frac{r_-}{r}
\right )^d\right ]^{\alpha d(d+1)/(d-1)},\\
F_{tr}&=&\bar{F}_{tr}=\frac{Q}{r^{d+1}},
\end{eqnarray}
where $d\bar{s}^2$ is given by Eq. (24). The charged black hole solution
 (37) has asymptotically flat region. The scalar field is bounded at the
  horizon, vanishes at  singular point $r=r_-$,  and tends to $\phi=1$ at
spacelike infinity. The action (8) and its equations of motion cannot 
remain unchanged under the transformation: $\bar{\phi}\rightarrow 
-\bar{\phi}$
if the Maxwell field is present. Therefore, unlike the absence of
 Maxwell field, we  obtain a  solution (37)-(39).
 In addition, it
is easy to show that the ADM mass, Hawking temperature, and the
 entropy of black hole (37) are still given by  (34)-(36), 
 respectively. This is so because the Euclidean action is invariant
under the conformal transformation (up to a surface term 
associated with the scalar field).  But it seems that the entropy
 of solution (24) satisfies the area formula, but the entropy 
 in solution (37)  doesn't. 
This is due to the fact that the black hole entropy
comes from the surface term in the Euclidean action formalism.
 The surface term in the Einstein frame (8) is given by
\begin{equation}
\bar{I}_{\rm surface}=-\frac{1}{8\pi}\int _{\partial \bar{V}}d^{d+2}x
\sqrt{\bar{h}}[\bar{K}-\bar{K}_o],
\end{equation}
where $\bar{K}$ represents the extrinsic curvature in the metric
$\bar{h}$ of a constant $r>r_+$ timelike supersurface $\partial \bar{V}$. 
And  $\bar{K}_o$ is the extrinsic curvature of  vacuum
 background (here it is the D-dimensional Minkowski spacetime). One can
show that in Einstein frame the  entropy satisfies the
 $1/4$ area formula [19]
\begin{equation}
\bar{S}=-\frac{1}{8\pi}\int _{r_+}d^{d+2}x \sqrt{\bar{h}}[\bar{K}
-\bar{K_o}]
=\frac{1}{4}\bar{\Sigma}.
\end{equation}
Instead in the BD frame (1), the surface term leads to
\begin{equation}
I_{\rm surface}=-\frac{1}{8\pi}\int _{\partial V}d^{d+2}x \sqrt{h}
\phi [K-K_o].
\end{equation}
The black hole entropy in the Brans-Dicke frame is found to be
\begin{equation}
S=-\frac{1}{8\pi}\int _{r_+}d^{d+2}x\sqrt{h}\phi[K-K_o]
=\frac{1}{4}\phi (r_+)\Sigma,
\end{equation}
where $\Sigma$ is the area of horizon in the Brans-Dicke frame (37).
It appears that due to the scalar field, the area formula is no longer
 valid in the BD theory [4]. But making use of (37), it is found that 
  (41) is equal to (43)
 and the  entropy remains unchanged under the 
 conformal transformations.

 For the Hawking temperature (35), in the Einstein frame (24), 
 it can be calculated as
\begin{equation}
\bar{T}=\left. \frac{(A^2)'}{4\pi\sqrt{A^2B^2}}\right |_{r_+},
\end{equation}
where a prime denotes derivative with respect to $r$. In the
Brans-Dicke frame (37), it is
\begin{equation}
T=\left. \frac{(\Omega ^2 A^2)'}{4\pi \Omega ^2\sqrt{A^2B^2}}
\right |_{r_+}.
\end{equation}
Because the conformal parameter $\Omega ^2$ is regular at the horizon,
one can find that $\bar{T}$ is equal to $T$. Therefore, the
Hawking
temperature is an invariant quantity under conformal transformations 
only if the transformations are regular at event horizon.

The invariance of the ADM mass of black holes can be deduced  from
the first law of thermodynamics
\begin{equation}
dM=TdS + \cdots,
\end{equation}
where $\cdots$ means the work terms. Because  the Hawking
temperature and  entropy are  invariant quantities, the ADM mass 
 must be invariant under the regular conformal transformations.

The thermodynamics of the  black hole solution (37) is quite 
interesting. When $d=1$ ($\alpha =0 $), it reduces to  that of
Reissner-Nordstr\"{o}m black holes (see the discussion below). 
For $d>1$ and  $r_+=r_-$, it follows from  (35) and
(36) that the Hawking temperature diverges because of $\alpha
 (d+1)/2>1, \bar S=0$.
This property is similar to  that  of the
$a>1$ dilaton black holes [20].

As for the  solution (37) in the Brans-Dicke-Maxwell
theory, we have three points to be stressed. First, when $D=4$,
 the solution (37)-(39) reduces to
 the Reissner-Nordstr\"{o}m solution with the constant scalar field
 ($\phi =1$). That is, in four dimensions, the 
 black hole solution
 is just the Reissner-Nordstr\"{o}m solution. 
Because the equation of motion for  scalar 
becomes source-free equation [see  (4) or (11)], there is nothing
 to support the nontrivial scalar field. Accordingly, we have only
the trivial scalar. This reason also holds for the vacuum BD
 theory. Second, in higher dimensions 
($b\ne 0$), we have the  black hole solution with the nontrivial
 scalar field in the BD-Maxwell
 theory. As is well known, the Maxwell field is allowed by the no-hair
 theorem of black holes. From  (4) or (11), we see that
the Maxwell field can be considered as the source of the scalar field.
 In this way the Maxwell field supports the nontrivial scalar field.
Here we remind the reader  that the stress-energy
tensor of the Maxwell field is traceless only in the four dimensions.
 Also in the Einstein frame, the action of the Maxwell
field is  invariant under  four-dimensional conformal transformations.
 This can be seen clearly from the action
 (8). Finally, we make a comment on the no-hair theorem.
At present, the interpretations of no-hair theorem seem to have
two aspects. One group argues that
an asymptotically flat black hole cannot carry the nontrivial scalar
 fields bounded on the regular horizon (see
the recent proofs of no-scalar hair theorem [11,12,21,22]. On the other
 hand,
 the no-hair theorem means that the black hole can be characterized by
 only a few parameters: mass, angular momentum, and electric (magnetic)
charge. In the latter sense, the existence of nontrivial scalar field in
 the BD-Maxwell theory does not violate the no-hair theorem because
the solution (37) depends  on only  two parameters: the ADM mass 
and electric
 charge. In fact, such scalar hairs
 exist largely in the black holes of string theories. 

\begin{flushleft}
{\bf \large Acknowledgments}
\end{flushleft}

The research of R.G.C. was supported in part by China Postdoctoral
 Science Foundation. He would like to thank Professor Y. Z. Zhang 
for helpful discussions.

\end{document}